\documentclass[12pt,preprint]{aastex}
\usepackage{amsmath,amsfonts}
\begin{document}
\title{Precision radial velocities of double-lined spectroscopic binaries
with an iodine absorption cell}
\author{Maciej Konacki \altaffilmark{1}}
\affil{Department of Geological and Planetary Sciences, California
Institute of Technology, MS 150-21, Pasadena, CA 91125}
\altaffiltext{1}{e-mail: maciej@gps.caltech.edu}
\begin{abstract}
A spectroscopic technique employing an iodine absorption cell (I$_2$) to 
superimpose a reference spectrum onto a stellar spectrum is currently the
most widely adopted approach to obtain precision radial velocities of
solar-type stars. It has been used to detect $\sim$80 extrasolar planets 
out of $\sim$130 known. Yet in its original version, it only allows us to 
measure precise radial velocities of single stars. In this paper, we present 
a novel method employing an I$_2$ absorption cell that enables us to accurately 
determine radial velocities of both components of double-lined binaries. Our 
preliminary results based on the data from the Keck~I telescope and HIRES 
spectrograph demonstrate that 20-30~ms$^{-1}$ radial velocity precision can 
be routinely obtained for ``early" type binaries (F3-F8). For later type
binaries, the precision reaches $\sim$10~ms$^{-1}$. We discuss applications of the 
technique to stellar astronomy and searches for extrasolar planets in binary 
systems. In particular, we combine the interferometric data collected with 
the Palomar Testbed Interferometer with our preliminary precision velocities 
of the spectroscopic double-lined binary HD~4676 to demonstrate 
that with such a combination one can routinely obtain masses of 
the binary components accurate at least at the level of 1.0$\%$. 
\end{abstract}


\section{Introduction}

The idea of passing starlight through an absorption medium to superimpose
a set of reference lines and subsequently precisely measure radial velocities 
was first proposed by \cite{Grif:73::}. At that time, \cite{Grif:73::} used the
atmosphere and the telluric H$_{2}$O and O$_{2}$ lines. In order
to overcome systematic errors in velocities caused by, among others, the
atmospheric pressure changes and hence shifts in the position of the
telluric lines, \cite{Cam:79::} introduced for the first time an absorption
cell (containing a toxic hydrogen fluoride gas). This technique was 
subsequently used to carry out a pioneering precision radial velocity survey 
of 16 stars over the period of six years \cite[1981-1987;][]{Cam:88::}. 
It is conceivable that \citeauthor{Cam:88::} would have detected the first 
extrasolar planet if they targeted a larger number of stars as their average 
velocity precision was $\sim$13~ms$^{-1}$.

\cite{Mar:92::} took the absorption cell technique to a new level by
introducing an iodine (I$_{2}$) absorption cell with the modeling
of a spectrograph point spread function (PSF). The gaseous iodine, besides being much less toxic,
has a strong line absorption coefficient (and hence requires a path length
of only a few centimeters) and offers a dense forest of absorption lines over
the wavelength range of 500-630 nm. In their approach, the Doppler shift
of a star spectrum $\Delta\lambda$ is determined by solving the following 
equation \citep{Mar:92::}
\begin{equation}
\label{i2::}
I_{obs}(\lambda) =
[I_{s}(\lambda+\Delta\lambda_{s})\,T_{I_{2}}(\lambda+\Delta\lambda_{I_{2}})]
\,\otimes\,PSF
\end{equation}
where $\Delta\lambda_{s}$ is the shift of the star spectrum,
$\Delta\lambda_{I_{2}}$ is the shift of the iodine transmission 
function $T_{I_{2}}$, $\otimes$ represents a convolution and
$PSF$ a spectrograph PSF. The parameters $\Delta\lambda_{s},
\Delta\lambda_{I_{2}}$ as well as parameters describing the PSF
are determined by performing a least-squares fit to the observed
(through the iodine cell) spectrum $I_{obs}$. To this
end, one also needs a high SNR star spectrum taken without
the cell $I_{s}$ which serves as a template for all the
spectra observed through the cell and the I$_2$ transmission
function $T_{I_{2}}$ obtained with the Fourier
Transform Spectrometer at the Kitt Peak National Observatory.
The Doppler shift of a star spectrum is then given by
$\Delta\lambda = \Delta\lambda_{s} - \Delta\lambda_{I_{2}}$.

To date, the iodine technique has been successfully used to detect 
$\sim$80 out of $\sim$130 known extrasolar planets \citep{Sch:04::}. 
Most of these detections were accomplished by the California-Carnegie 
group \citep{Mar:03::}. The iodine technique --- thanks to its conceptual
simplicity --- is the most commonly adopted way 
to obtain precision radial velocities. Iodine absorption cells are 
available on many spectrographs --- HIRES at the 10m Keck I 
(Keck Observatory), Hamilton at the 3m Shane (Lick Observatory), 
UCLES at the 3.9m Anglo-Australian Telescope (Anglo-Australian Observatory),
HRS at the 9m HET (McDonald Observatory), MIKE at the 6.5m Magellan 
(Las Campanas Observatory), UVES at the 8m Kueyen (Cerro Paranal),
HDS at the 8.2m Subaru (National Astronomical Observatory of Japan)
and many others --- and are used for planet detections. 
Unfortunately, in its classical version the iodine technique can only 
be applied to single stars or spectroscopic binaries whose secondaries 
are so faint that their spectral lines are undetectable. This is dictated 
by the need to supply an observed template spectrum of each binary
component for Eq.~\eqref{i2::}, but in the case of double-lined spectroscopic 
binaries it cannot be accomplished since their spectra are obviously always
composite and time variable.

In this paper, we present a novel technique which employs an I$_2$
absorption cell to obtain precision radial velocities of both
components of double-lined spectroscopic binaries. The technique
and its tests are described in Section 2 and its applications to
stellar astronomy and searches for planets in binary and multiple
stellar systems are discussed in Section 3.

\section{The Method}

We can measure precise radial velocities of both components of a spectroscopic 
binary with an I$_2$ absorption cell in the following way. First, we always take 
two subsequent exposures of each (binary) target --- one with and the other without 
the I$_2$ cell. This is contrary to the standard approach for single stars where 
an exposure without the cell (a template) is taken only once. This way we obtain 
an instantaneous template which is used to model only the adjacent exposure taken 
with the cell. Next, we perform the usual least-squares fit and obtain the parameters
described in Eq.~\eqref{i2::}. Obviously, the derived Doppler shift,
$\Delta\lambda_i$ (where $i$ denotes the epoch of the observation), carries
no meaning since each time a different template is used (besides it
describes a Doppler ``shift" of a composite spectrum that is typically different at
each epoch). However, the parameters (in particular the wavelength solution and 
the parameters describing PSF) are accurately determined and can be used to extract 
the star spectrum, $I^{\star,i}_{obs}(\lambda)$, for each epoch $i$, by
inverting the equation~\eqref{i2::}
\begin{equation}
\label{met::}
I^{\star,i}_{obs}(\lambda) = [I^{i}_{obs}(\lambda)\,\otimes^{-1}\,PSF^{i}]
/T_{I_{2}}(\lambda)
\end{equation}
where $\otimes^{-1}$ denotes deconvolution \cite[carried out using a modified Jansson
technique;][]{Gil:92::} and $PSF^{i}$, symbolically, the set of parameters 
describing PSF at the epoch $i$. Such a star spectrum has
an accurate wavelength solution, is free of the I$_2$ lines and the influence 
of a varying PSF. In the final step, the velocities of both components
of a binary target can be measured with the well known two-dimensional
cross-correlation technique TODCOR \citep{Zuc:94::} using as templates the 
synthetic spectra derived with the ATLAS~9 and ATLAS~12 programs 
\citep{Kurucz:95::} and matched to the observed spectrum, \citep{Kon:03a::}
$I_{s}(\lambda)$. The formal errors of the velocities can be 
derived from the scatter between the velocities from different echelle 
orders or using the formalism of TODCOR \citep{Zuc:94::}.

The above approach will produce radial velocities that may suffer from two
possible sources of systematic errors. (1) The numerical procedure of
extracting a star spectrum from an exposure taken with the I$_2$ cell could
imaginably introduce systematic errors. (2) TODCOR itself could also produce
systematic errors. It has never been tested at a high velocity precision 
(let us say, $\sim$10~ms$^{-1}$) and the use of synthetic spectra could 
be a source of systematic errors. 

In order to demonstrate our method and investigate the systematic errors,
we have observed a number of targets with HIRES spectrograph at the Keck I
telescope. HIRES equipped with an I$_2$ absorption cell has been successfully 
used to detect many extrasolar planets \citep{Mar:03::}. We have observed 
HD~209458 (a single star with a known short period extrasolar planet, V=7.6 mag), 
HD~109358 (a velocity standard, V=4.3 mag), HD~4676 and HD~282975 (double-lined 
spectroscopic binaries in Pleiades, V=5.0 and 10.0 mag, respectively) and HD~206901 
(a triple system, V=4.2 mag). The data span is approximately two years for 
HD~209458 (from mid 2002 to mid 2004), one year for HD~4676, HD~282975 
and HD~206901 (from mid 2003 to mid 2004) and $\sim$10 
minutes for HD~109358 (on 29 Mar 2004). With the exception of HD~209458 
whose spectra had a typical SNR of 40-80 \citep[HD~209458 was initially used 
as a standard star for a separate transiting planet search and was observed 
with moderate exposure times; see][]{Kon:03a::,Kon:03b::}, the SNR of all 
other spectra was typically in the range of 150-300. 

Since in our approach we do not assume that an observed object must have
a composite spectrum, the method can be applied to single stars as well. 
Obviously, there is no need to analyze the spectra of single stars in 
such a way but the velocities computed in both ways (Eqs 1 and 2) can be 
compared and used to explore possible additional errors originating
in the proposed procedure. Two such comparisons are presented in 
Figures 1 and 2. The first figure shows 6 velocity measurements of HD~109358
over the period of about 10 minutes. HD~109358 is a high-precision 
radial velocity standard proposed by the Geneva group \citep{Ud:99::}.
Clearly, the two velocity sets have very similar scatter and no additional
errors at the 3 ms$^{-1}$ can be seen. The second test (Fig.~2) concerns
HD~209458 (the data span is two years), a well known star harboring a transiting giant 
planet with the orbital period of 3.52 days. Its orbital parameters are
very accurately known \citep{Na:04::} and hence were fixed in this test. 
The only parameter for which we fit was a velocity offset. The test 
does not reveal any additional errors either. In fact, the velocities 
computed using a cross-correlation function (CCF) have a smaller rms 
(10 vs 14 ms$^{-1}$) than the velocities computed with the traditional 
iodine approach --- most likely due to a mediocre SNR of the spectra 
of HD~209458 (the iodine technique with PSF modeling works best on 
high SNR spectra).

In order to test our method on a truly binary spectrum we have observed 
HD~4676 (64 Psc) --- a nearby (23 pc) binary star composed of two F8 dwarfs. 
HD~4676AB has the orbital period of 13.8 days and the brightness ratio of $\sim 0.9$. 
\cite{Abt:76::} were the first to determine its spectroscopic orbit which
was later improved by \cite{Nad:79::} and \cite{Duq:91::}. Moreover, its astrometric 
orbit is known from the Palomar Testbed Interferometer \cite[PTI;][]{Bod:99::}.
This has important consequences for our work as such astrometric data
can be used to verify our precision radial velocities --- orbital models 
for astrometric and spectroscopic measurements share most of the orbital 
parameters and can be used together to perform a combined least-squares fit 
\cite[see e.g.][]{Kon:04::}. 

We have observed HD~4676 over the period of one year on five occasions
(11 Aug 2003, 12 Aug 2003, 17 Nov 2003, 18 Nov 2003 and 18 Jul 2004)
and obtained 24 radial velocity (RV) measurements for each component. The
average formal RV error is 25 ms$^{-1}$ for the primary and 28 ms$^{-1}$
for the secondary. The quality of the data was verified by performing a
combined fit to our RVs and the astrometric data published by \cite{Bod:99::}. 
The details of the fit are presented in Figure~3. The post-fit rms of 
24 ms$^{-1}$ is in a very reasonable agreement with the formal RV errors 
and is 27 times smaller than the rms reported by \cite{Duq:91::}.
It obviously results in a much better determination of the orbital and
physical parameters of the system. We discuss it in detail in Section 3.1.

The most important remaining issue is whether TODCOR and the use of synthetic 
spectra could introduce any systematic errors to RVs. While errors that simply 
increase the post-fit rms are not particularly troubling, one can in principle
expect errors that may depend on the orbital phase and possibly affect the
best-fit values of the orbital parameters. A common sense way to detect
such errors is to apply our method to simulated spectra of HD~4676 and try
to recover the RVs assumed in the simulation. We have performed such tests
and have not detected any systematic errors. There is however an even
better way to control these errors. As already mentioned, thanks to a high 
precision astrometry of HD~4676 from PTI \citep{Bod:99::}, one can perform 
a combined fit to the astrometric and spectroscopic data. If there existed
any significant phase-dependent systematic variations of the velocities,
they would be detectable in the post-fit residuals from the combined solution. 
Clearly, these are absent in our solution (Fig.~3) and $\chi^2/DOF$ is $1.18$.

It should be noted that the best-fit rms of 24 ms$^{-1}$ for the RVs of
the primary and secondary is clearly larger than one would expect if the
RV errors were photon-limited \cite[2-3 ms$^{-1}$;][]{But:96::}. Possible
sources of additional errors include: a mismatch between the synthetic
templates and observed spectra and astrophysical phenomena 
\cite[e.g. starspots, convective inhomogeneities, see][]{Saa:98::}.
We do not have yet enough data for other binaries to fully address
this issue. Nevertheless, it is reasonable to expect that for later
type stars (with more spectral lines), the velocity precision will be
higher. This appears to be confirmed by our velocities of HD~282975, 
a double-lined spectroscopic binary from Pleiades ($P_{orb}=26^{d}$), 
which is composed of two G6 dwarfs. The post-fit rms is 11 ms$^{-1}$ for 
the primary and 13 ms$^{-1}$ for the secondary and is $\sim 80$ times
smaller than the rms reported by \cite{Mer:92::}. The best fit is based
on only 8 RVs but from our experience with HD~4676 we know that 
early accuracy is later confirmed by a larger data set.

Finally, let us note that the iodine technique for single stars typically
produces RVs which have an arbitrary offset \cite[unless special steps are
undertaken, see][]{Nid:02::}. In our approach which employs synthetic spectra 
one can in principle calibrate and refer the velocities to a standard system of 
stars. We plan to do it once we collect enough observations of the RV standards.

\section{Applications}

\subsection{Stellar astronomy}

Modern tests of stellar structure and evolution models require stellar
masses accurate to $2\%$ or better \citep{And:91::}. Stellar masses are
traditionally derived from the observations of binary stars by combining 
(1) photometric (light curves) and spectroscopic data (RVs) for double-lined 
eclipsing binaries, (2) relative astrometry and spectroscopic data for 
double-lined binaries, (3) from spectroscopy, relative astrometry and 
parallax for single-lined binaries and (4) absolute astrometry. There are 
currently known about $60$ binaries with masses of the components accurate 
at the level of $2\%$ or better \cite[see e.g.][]{Las:02::}. Almost all 
of them are eclipsing binaries and most of them have components with the 
masses in the 1-3 M$_{\odot}$ range (more massive and hence larger stars
are more likely to be eclipsing, for a given orbital period).

Optical interferometry offers the unique opportunity to determine visual
orbits of double-lined spectroscopic binaries \cite[apparent semi-major 
axes at the  milliarcsecond level; for a recent discussion see][]{Kon:04::} 
and subsequently masses of the components, regardless of their size. Up 
to date, about twenty spectroscopic binaries have had their orbits determined 
with optical interferometers \citep{Qui:01::}. In a few cases, when combined 
with radial velocity measurements, they already produced component masses 
accurate at the level of 1-2$\%$.

Our RVs constitute an excellent complement to precise astrometric 
orbits from optical interferometers. This is demonstrated well
by HD~4676. \cite{Bod:99::} combined their 
interferometric data of HD~4676 from PTI with the radial velocity 
measurements of \cite{Duq:91::}. They have determined the masses
to be $M_{1} = 1.223\pm0.021$M$_{\odot}$ (1.7$\%$), $M_{2} =
1.170\pm0.018$M$_{\odot}$ (1.5$\%$), the velocity amplitudes 
$K_{1} = 57.35\pm0.31$ kms$^{-1}$, $K_{2} = 59.95\pm0.32$ kms$^{-1}$
and the inclination, $i = 73.80\pm0.92$ deg. The relative error in the 
masses ($\Delta M/M$) is dominated by the error in the orbital
inclination, $3 \Delta i\cos(i)/\sin(i) \approx 1.4\%$, but the errors
in the velocity amplitudes also contribute. In order to make the best
out of PTI data, we used our RVs and performed a new combined fit.
The best-fit orbital and physical parameters of the system
are presented in Tables~1 and 2. The relative error in the masses
is now $1.2\%$ and the overall quality of the fit has improved. 
In this specific case, the ultimate accuracy of the masses is still 
limited by the accuracy of the orbital inclination. However, since
the radial velocities are so precise, one can simply take more 
astrometric data with PTI and improve the masses even more.
Clearly, our approach will be of great use for all those binaries
that have orbital inclinations closer to an edge-on configuration.
In particular, light curves of eclipsing binaries combined with
precise RVs should produce masses accurate at the $0.1\%$ 
level on a regular basis. 

Precise RVs can obviously be used not only to determine the masses of 
the companions but the distance to the binary as well. For example, the 
parallax of HD4676 is formally very accurate, $43.496\pm0.089$ mas 
(the distance of $22.991\pm0.047$~pc; the final accuracy will depend on how 
well the brightness ratio of the binary components is known). In this 
context, HD~282975 (Fig.~4) appears as a particularly interesting binary 
as it can be used to derive a very precise distance to Pleiades.
To this end, one needs interferometric-astrometry
of HD~282975 which should be easy to obtain with the Keck Interferometer 
(HD~282975 is too faint for PTI). Finally, precise RVs can be used to explore
the regime of circular or almost circular orbits (the error of the
eccentricity of HD~4676 is only $6.0\times10^{-4}$, Tab.~1) and
the cut-off period below which orbits are circularized.

\subsection{Extrasolar planets in binary stellar systems}

\cite{Duq:91::} have demonstrated that the frequency of 
binaries (BF) among field stars older than 1 Gyr is 57\%. The studies 
of multiplicity of pre main-sequence stars (PMS) in the Taurus and
Ophiuchus star forming regions show that BF for systems in
the separation range 1 to 150 AU is twice as large as that of the
older field stars \citep{Sim:95::}. Further investigations have
concluded that BF is lower for young stellar clusters (and similar
to BF of the field stars) and that the binary frequency for PMS
seems to be anti-correlated with the stellar density \citep{Mat:00::}.
Nonetheless, BF is very high for both field and pre main-sequence
stars and one can argue that it may not be possible to assess 
the overall frequency and properties of extrasolar planets 
without addressing binary (and multiple) stellar systems. 

Supporting arguments come from the presence of circumstellar and 
circumbinary disks around binaries. One of the prime examples of 
the circumstellar disks in a binary system is the case of L1551~IRS~5. 
\cite{Rod:98::} show that L1551~IRS~5 is a binary PMS with the separation 
of 45 AU in which each component is surrounded by a disk. The radii 
of the disks are 10 AU and the estimated masses are 0.06 and 
0.03 $M_{\odot}$, supposedly enough to produce planets. Recently, 
\cite{Mcc:03::} have spatially resolved mid-infrared scattered light from the
protoplanetary disk around the secondary of the PMS binary
HK Tau AB. The inferred sizes of the dust grains are in the range
1.5-3 $\mu$m which suggests that the first step in the planet
formation, the dust grain growth, has occurred in this disk.
Millimeter and submillimeter
measurements of the dust continuum emission enable us to measure the
total masses of disks. These observations show that the circumbinary disks
may be reduced in size and mass but still are present even in
close systems. The circumbinary disks are observed at
millimeter wavelengths around many PMS spectroscopic binaries.
Such massive disks are however rare around wide binaries with
separations of 1-100 AU. This is confirmed by numerical works
that predict circumstellar and circumbinary disks
truncated by the companions \citep{Lub:00::}. The circumstellar
disks have the outer radii 0.2-0.5 times the binary separation while
the circumbinary disks have the inner radii 2-3 times the
semi-major axis of the binary. Finally, the measurements of the
infrared excess emission show no difference in frequency of the
excess among binaries and single stars. It indicates that the
circumstellar material in binary systems may be similar in
temperature and surface density to that in the disks surrounding
single stars \citep{Mat:00::}. 

The problem of stability of the planetary orbits in binaries 
was recognized a long time ago. It was mostly approached with
the numerical studies of the elliptic restricted three-body
problem. The orbital configurations considered include the
so-called P-type (Planet-type, circumbinary orbits), S-type
(Satellite-type, circumprimary or circumsecondary orbits) and
L-type orbits (Librator-type, orbits around stable Lagrangian
points L4 or L5 for the mass ratios $\mu<0.04$). There are many
papers concerning the stability of S-type motions
\cite[e.g.][]{Ben:03::,Pil:02::,Rab:88::}. They aimed at 
developing empirical stability criteria in the framework of the 
circular three-body problem \cite[see e.g.][]{Gra:81::,Bla:82::,
Pen:83::}, the stability of periodic orbits and the stability 
of the test particles in binary systems as a function of the 
eccentricity of the binary. The P-type motions were also investigated 
\cite[]{Pil:03::,Bro:01::,Hol:99::}; as were the L-type 
orbits \citep[see e.g.][]{Lau:02::}. Unfortunately, most of the 
analytical papers deal only with circular binary orbits,
numerical studies are confined to special mass ratios 
and the integration times used are relatively short. Also, they are 
almost exclusively restricted to the framework of the three body 
problem. Some of these issues are addressed by \cite{Hol:99::} 
who studied a range of mass ratios, eccentricities and used long 
integration times (at least $10^4$ periods of the binary). 
Nonetheless, we can safely conclude that planets in binary stellar 
systems, once formed, would enjoy a wide range of stable orbits.

However, the theories of planet formation in binary stellar systems are
still at early stages. \cite{Whi:98::} studied terrestrial planet growth
in the circumprimary habitable zones in a binary system. They
considered a 4-body system of 2 stars and 2 planetesimals for
which by varying the orbital parameters of the binary (the semi-major axis, 
eccentricity and mass ratio) they were able to determine a critical 
semi-major axis of the binary below which the secondary does not allow a
growth of the planetesimals (planetesimals are accelerated by the
secondary, their relative velocity is larger than
critical and their collisions become destructive). Based on this
criterion, they concluded that about 60\% of nearby solar-type
binaries cannot be excluded from having a habitable planet.
\cite{Mar:00::} analyzed $\alpha$~Cen~AB (the semi-major axis of 23 AU,
eccentricity of 0.52, mass ratio 1.1/0.92), a prototype close
binary system, and demonstrated that the planetesimals can accrete
into planetary embryos. \cite{Bar:02::} continued the study and showed
that planetary embryos can grow into terrestrial planets in about
50 Myr. Somewhat contrary to this result, \cite{Nel:00::}, who analyzed a
binary system similar to L1551~IRS~5, found that the planet
formation is unlikely in equal mass binary systems with the separation
of about 50 AU. Yet another result by \cite{Bos:98::} claims that a
stellar companion can induce a multi-Jupiter-mass planet
formation. Clearly, there is a lack of consensus and the planet
formation theories would certainly benefit from observational
constraints.

Unfortunately, current RV searches for extrasolar planets have been
limited to single stars or binaries with large apparent separations.
Despite that prejudice out of $\sim$130 extrasolar planets, eighteen belong 
to stars which have also stellar companions \citep{Egg:04::}. They even seem 
to posses characteristics different from those orbiting single stars. As recently 
pointed out by \cite{Egg:04::}, (1) the most massive short period planets
orbit stars from multiple stellar systems and (2) all these planets
($P_{orb} < 40$ days) tend to have very low eccentricities. It is a
surprisingly interesting result given that current RV surveys suffer from
a selection effect --- the binary stars with separations smaller
than $\sim$2 arcsec have been excluded from the surveys. The exception 
here is HD~41004AB which has a separation of 0.5 arcsec. Presumably, 
this star entered the Geneva RV program as a single star and was later found 
to be a triple (sic!) system in which the primary has apparently a giant 
planet \cite[]{Egg:04::} and the secondary and tertiary are very faint
(an M star and a putative brown dwarf). Thus these stars from binary 
or multiple stellar systems which we know that harbor planets 
have a distant or/and a faint stellar 
companion.  There is a need to perform an unbiased survey for extrasolar 
planets and substellar companions to stars from multiple systems which will probe
different physical separations and brightness ratios. 

To this end, two years ago we have initiated a survey to detect planets in 
binary stellar systems with the Keck~I and HET telescopes. Our survey 
covers about $\sim$450 binary stars from the northern and southern hemispheres. 
Preliminary results for one of our representative targets, $\kappa$~Peg are shown 
in Fig.~5. $\kappa$~Peg (HR~8315, HD~206901, HIP~107354) is a known triple system 
of which the spectral lines of the two brightest components AB (apparent
semi-major axis 0.4 arcsec) can be easily
identified ($\Delta m = 0.04$, $P_{orb} = 11.59\;\mbox{yr}$) and the component 
B is itself a single-lined spectroscopic binary with the orbital period of 
$\sim$5.97 days \cite[e.g.][]{May:87::}. The radial velocities of the primary 
do not reveal any interesting variations and are of inferior quality due to
a large $v\sin i \approx 30\;\mbox{kms$^{-1}$}$. On the other hand, the RVs of the
sharp-lined secondary are more than acceptable (given an early spectral type of 
the star, F3IV) with the rms of $\sim 30\;\mbox{ms$^{-1}$}$. It is an improvement 
of 30 times in the rms compared to the work of \cite{May:87::} and 
demonstrates the usefulness of our approach. Since the iodine cells are
available on many high-resolution spectrographs, our method clearly 
opens exciting opportunities in the studies of other multiple systems.

\section{Summary}

We have introduced a new method to measure precision radial velocities
of double-lined spectroscopic binaries with an iodine absorption
cell. Our initial data sets demonstrate 20-30 ms$^{-1}$ velocity 
precision for ``early" (F3-F8) type and $\sim$10~ms$^{-1}$ for later
type binaries. If combined with interferometric astrometry, such
velocities enable us to determine the masses of the components of 
binary stars accurate at least at the level of $1\%$. Our method also
makes it possible to perform a search for extrasolar planets in
binary and multiple stellar systems where the brightness of the
primary and the secondary is comparable. This allows us to probe
a regime of the parameter space of the formation and evolution
of extrasolar planets not covered by other surveys.

\acknowledgements

M.K. would like to thank Shri Kulkarni for a very generous support and
guidance; Guillermo Torres and Dimitar Sasselov for many discussions on 
stellar spectroscopy; Eliza Miller-Ricci and Dimitar Sasselov for their 
high resolution grid of synthetic stellar spectra; Krzystof Gozdziewski 
for a discussion on the stability of planets in binary systems and the
referee and Matthew Muterspaugh for careful reading of the manuscript
and many useful suggestions.  The data presented herein were obtained at 
the W.\ M.\ Keck Observatory, which is operated as a scientific partnership 
among the California Institute of Technology, the University of California 
and the National Aeronautics and Space Administration. The Observatory was 
made possible by the generous financial support of the W.\ M.\ Keck Foundation. 
This work was supported by NASA through grant NNG04GM62G.

\clearpage

%
%

\figcaption[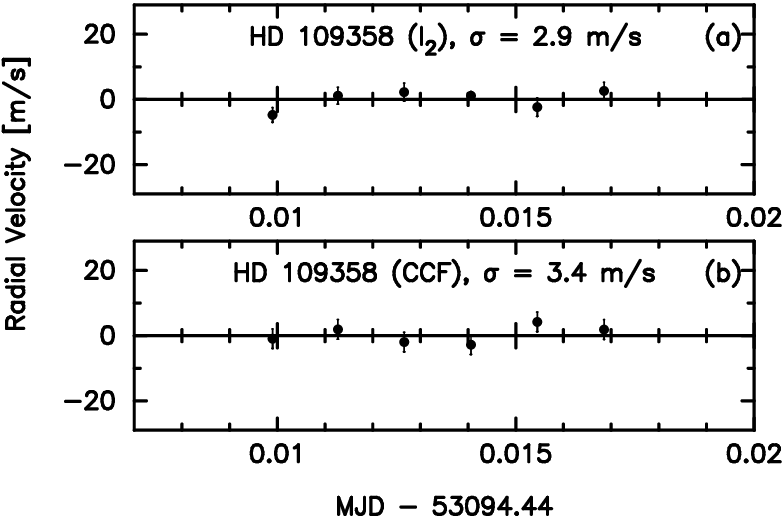]{Radial velocities of HD~109358 computed with ({\it a}) 
the traditional iodine (I$_2$) technique and with ({\it b}) the proposed method.}

\figcaption[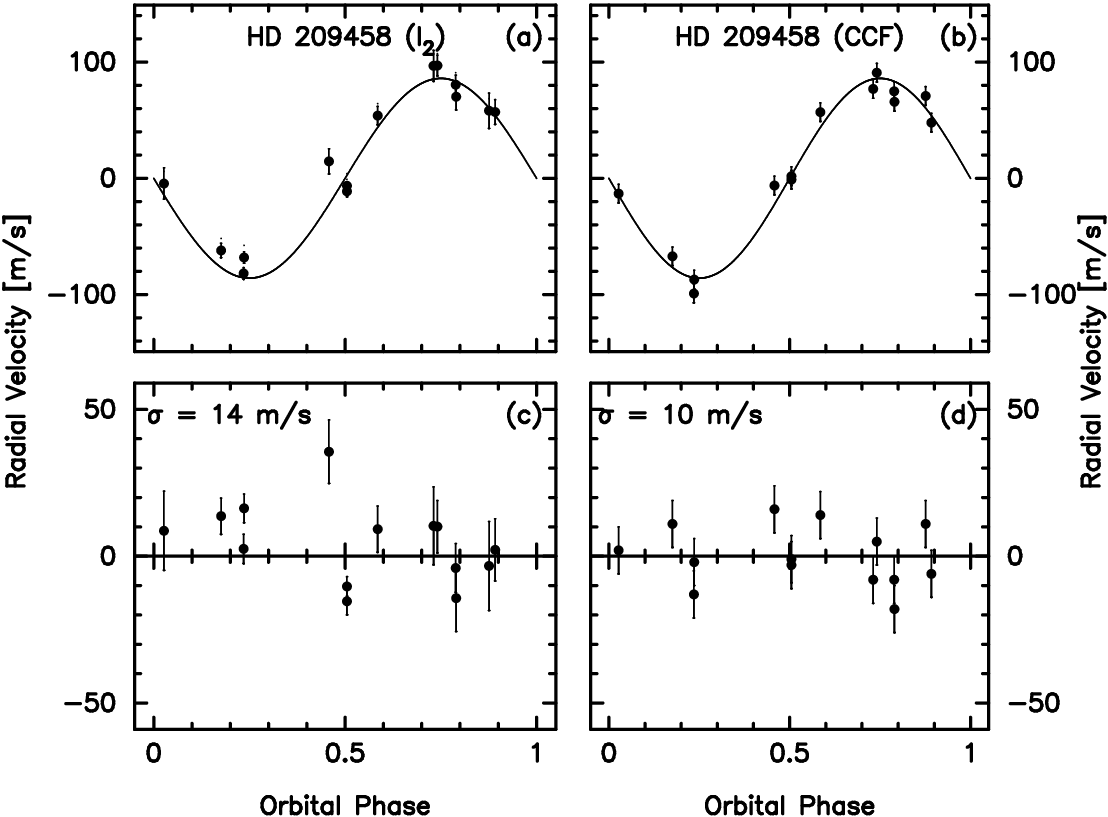]{Radial velocities of HD~209458 computed with ({\it a})
the traditional iodine (I$_2$) technique and with ({\it b}) the proposed method.
The data span is two years. ({\it c,d}) Best-fit residuals as a function of phase.}

\figcaption[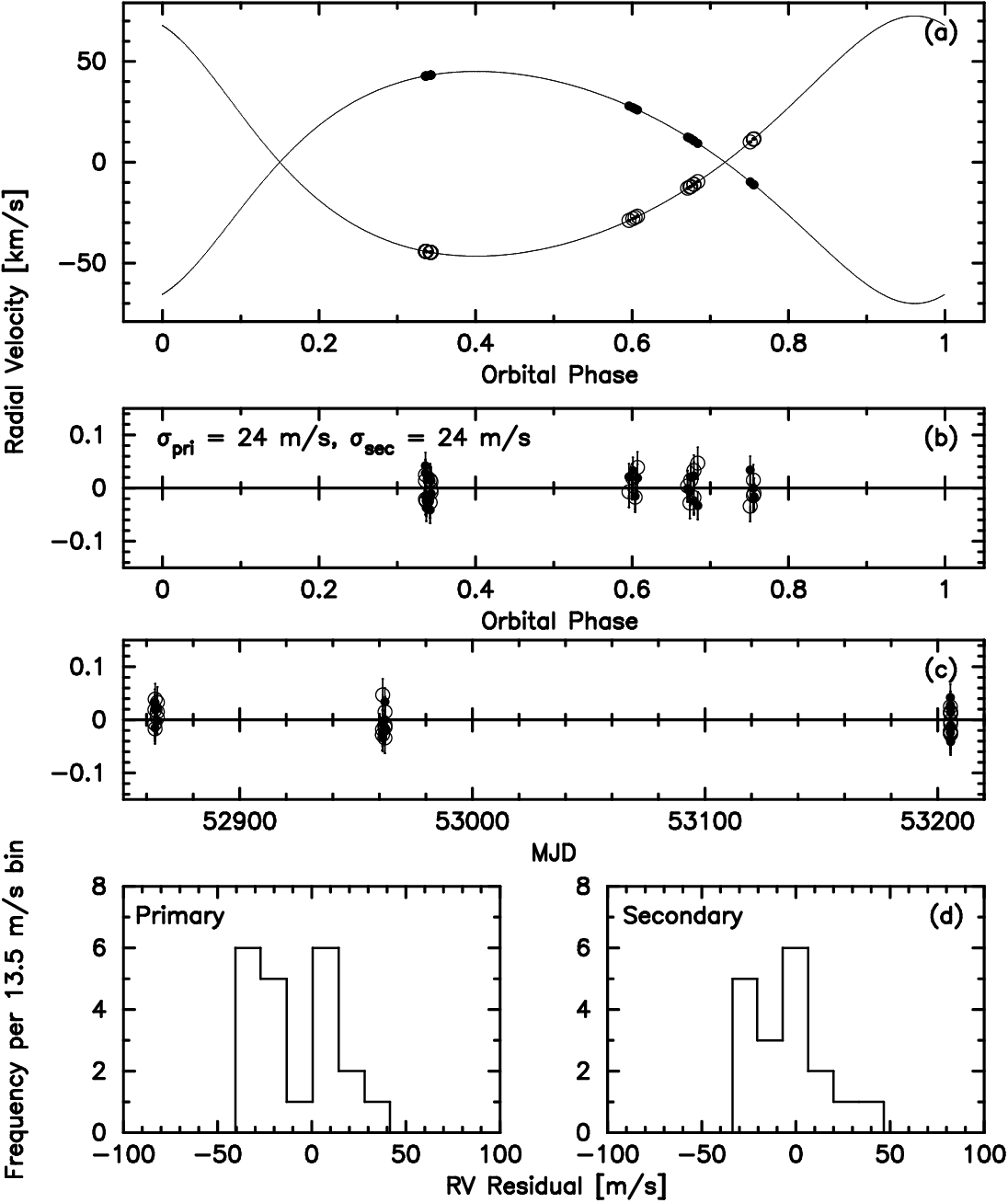]{Observed (filled circles for the primary and open
circles for the secondary) and modeled (solid line) radial velocities of
HD~4676. ({\it b,c}) Best-fit residuals as a function of phase and time. 
({\it d}) Histogram of the residuals.}

\figcaption[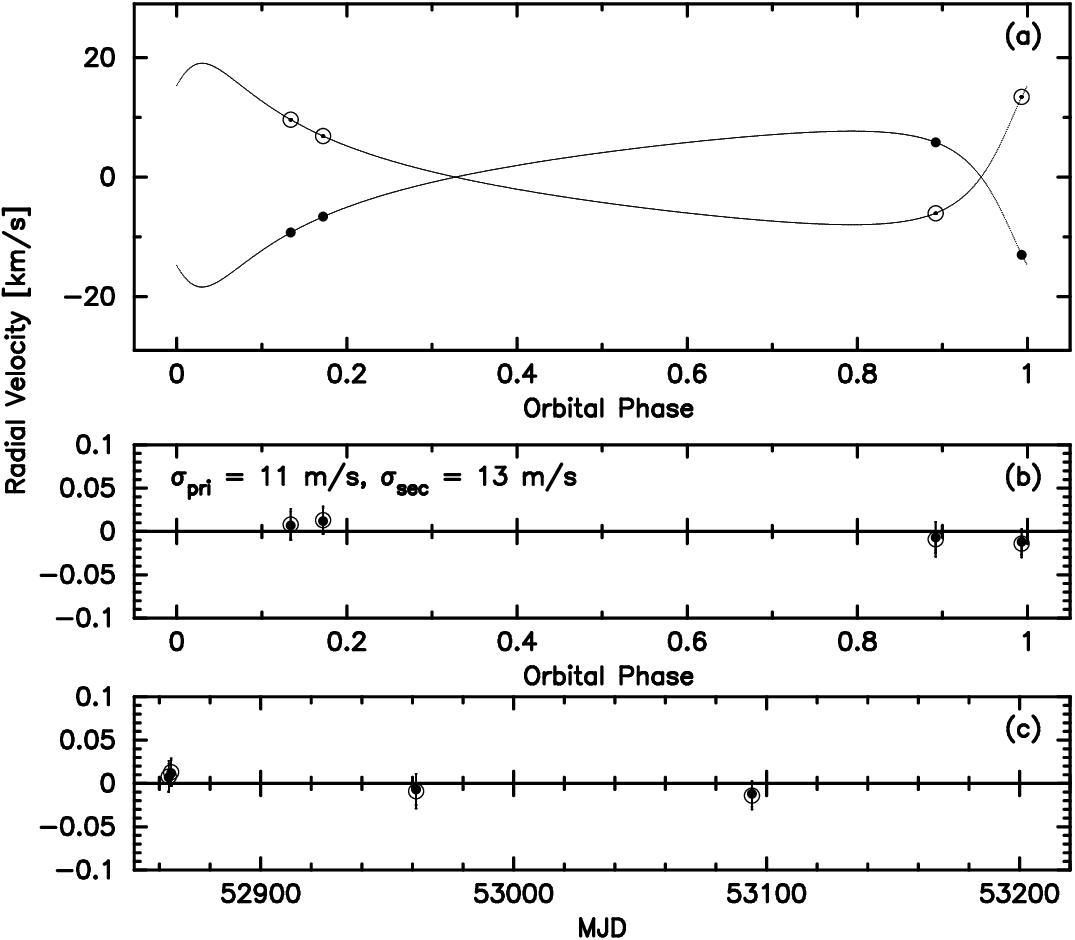]{Observed (filled circles for the primary and open
circles for the secondary) and modeled (solid line) radial velocities
of HD~282975. ({\it b,c}) Best-fit residuals as a function of phase and
time.}

\figcaption[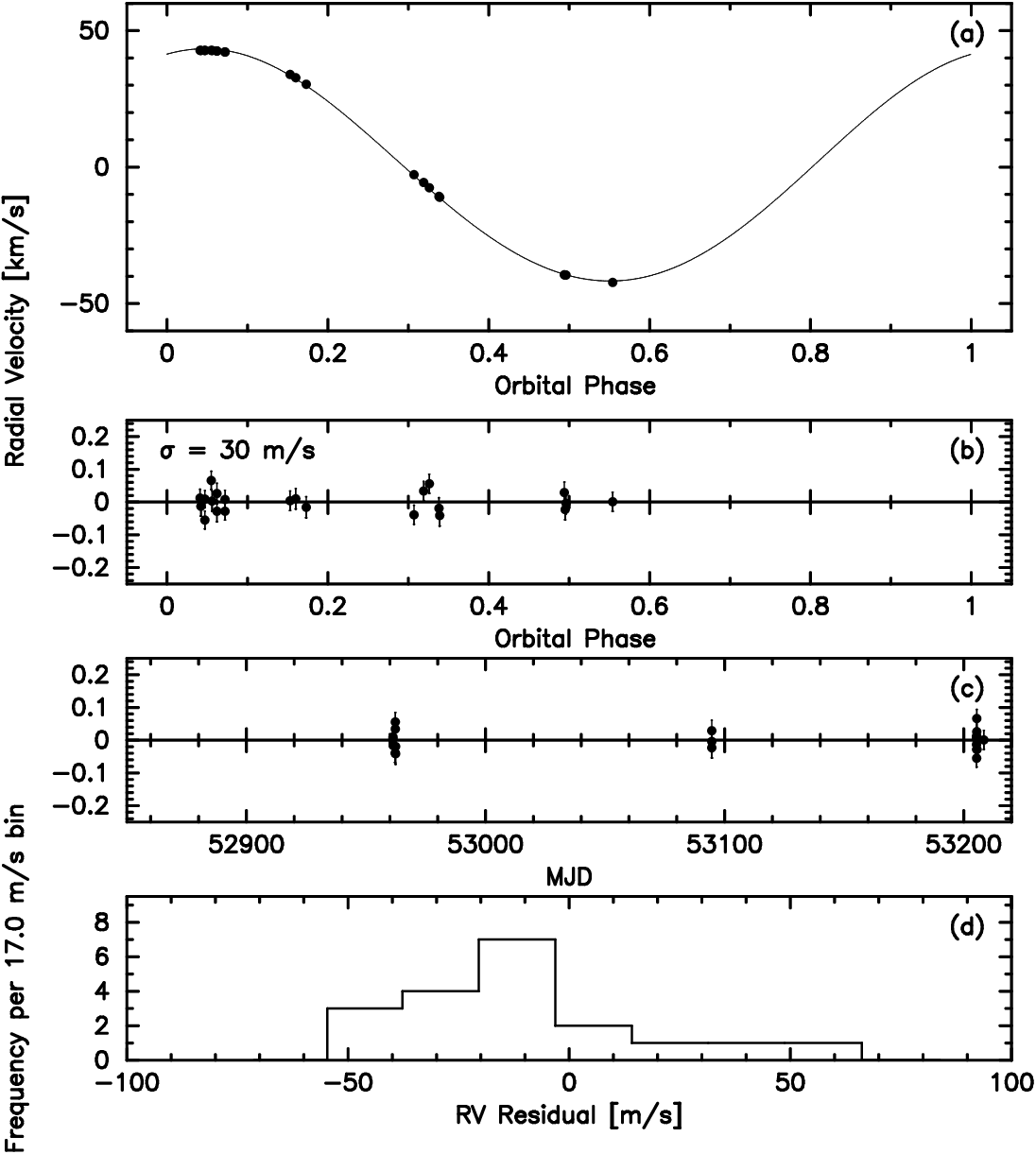]{Observed (filled circles) and modeled (solid line) 
radial velocities of HD~206901B. ({\it b,c}) Best-fit residuals as 
a function of phase and time. ({\it d}) Histogram of the residuals.}

\clearpage

%
%

%
%

\begin{figure}
\figurenum{1}
\epsscale{0.7}
\plotone{f1.eps}
\caption{}
\end{figure}

%
%

\begin{figure}
\figurenum{2}
\epsscale{0.8}
\plotone{f2.eps}
\caption{}
\end{figure}

%
%

\begin{figure}
\figurenum{3}
\epsscale{0.8}
\plotone{f3.eps}
\caption{}
\end{figure}

%
%

\begin{figure}
\figurenum{4}
\epsscale{0.8}
\plotone{f4.eps}
\caption{}
\end{figure}

%
%

\begin{figure}
\figurenum{5}
\epsscale{0.8}
\plotone{f5.eps}
\caption{}
\end{figure}

\clearpage
 
%
%

%
%

\begin{deluxetable}{lc}
\tablewidth{400pt}
\tablecaption{Best-fit Orbital Parameters for HD~4676\tablenotemark{a}}
\tablehead{\colhead{Parameter} & \colhead{HD~4676} }
\startdata
Apparent semi-major axis, $\hat{a}$ (mas) \dotfill & 6.545 $\pm$ 0.0133\\
Period, $P$ (d) \dotfill &  13.8244906 $\pm$ 4.3$\times 10^{-5}$\\
Time of periastron, $T_p$ (MJD) \dotfill & 50905.9746 $\pm$ 0.0067\\
Eccentricity, $e$ \dotfill &  0.23657 $\pm$ 0.00063\\
Longitude of the periastron, $\omega$ (deg) \dotfill & 203.057 $\pm$ 0.073\\
Longitude of the ascending node, $\Omega$ (deg) \dotfill &  207.41 $\pm$ 0.65\\
Inclination, $i$ (deg) \dotfill & 73.92 $\pm$ 0.80\\
Magnitude difference (K band, {\it assumed}), $\Delta K$ \dotfill & {\it 0.11} \\
Velocity amplitude of the primary, $K_1$ (km/s) \dotfill & 57.552 $\pm$ 0.037 \\
Velocity amplitude of the secondary, $K_2$ (km/s) \dotfill & 59.557 $\pm$ 0.038 \\
Reduced $\chi^2$, $\chi^2/DOF$ \dotfill & 1.18\\
\enddata
\tablenotetext{a}{Figures in parentheses are the $1\sigma$ formal errors}
\end{deluxetable}

%
%

\begin{deluxetable}{lcc}
\tabletypesize{\small}
\tablewidth{370pt}
\tablecaption{Physical Parameters for HD~4676\tablenotemark{a}}
\tablehead{\colhead{Parameter} & \colhead{Primary} &  \colhead{Secondary}}
\startdata
Semi-major axis, $a_{1,2}$ (AU)\dotfill  & 0.073953 $\pm$ 4.8$\times10^{-5}$ 
& 0.076529 $\pm$ 5.0$\times10^{-5}$ \\
Mass, $M$ (M$_{\odot}$)\dotfill   &  1.210 $\pm$ 0.014 & 1.169 $\pm$ 0.014 \\
Parallax, $\kappa$ (mas)\dotfill &  & \hspace{-5cm}{\mbox{$43.496 \pm 0.089$}} \\
Distance, $d$ (pc)\dotfill & & \hspace{-5cm}{\mbox{$22.991 \pm 0.047$}} \\
Spectral type \dotfill  & F8V & F8V \\
\enddata
\tablenotetext{a}{Figures in parentheses are the $1\sigma$ formal errors}
\end{deluxetable}


\begin{thebibliography}{}

\bibitem[Abt \& Levy(1976)]{Abt:76::} Abt, H.~A.~\& Levy, S.~G.\ 
1976, \apjs, 30, 273

\bibitem[Andersen(1991)]{And:91::} Andersen, J.\ 1991, \aapr, 3, 
91

\bibitem[Barbieri, Marzari, \& Scholl(2002)]{Bar:02::} Barbieri, 
M., Marzari, F., \& Scholl, H.\ 2002, \aap, 396, 219

\bibitem[Benest(2003)]{Ben:03::} Benest, D.\ 2003, \aap, 400, 
1103

\bibitem[Black(1982)]{Bla:82::} Black, D.~C.\ 1982, \aj, 87, 
1333

\bibitem[Boden et al.(1999)]{Bod:99::} Boden, A.~F., et al.\ 
1999, \apj, 527, 360

\bibitem[Boss(1998)]{Bos:98::} Boss, A.~P.\ 1998, Bulletin of 
the American Astronomical Society, 30, 1057

\bibitem[Broucke(2001)]{Bro:01::} Broucke, R.~A.\ 2001, 
Celestial Mechanics and Dynamical Astronomy, 81, 321

\bibitem[Butler et al.(1996)]{But:96::} Butler, R.~P., Marcy, 
G.~W., Williams, E., McCarthy, C., Dosanjh, P., \& Vogt, S.~S.\ 1996, 
\pasp, 108, 500

\bibitem[Campbell, Walker, \& Yang(1988)]{Cam:88::} Campbell, 
B., Walker, G.~A.~H., \& Yang, S.\ 1988, \apj, 331, 902

\bibitem[Campbell \& Walker(1979)]{Cam:79::} Campbell, B.~\& 
Walker, G.~A.~H.\ 1979, \pasp, 91, 540

\bibitem[Duquennoy \& Mayor(1991)]{Duq:91::} Duquennoy, A.~\& 
Mayor, M.\ 1991, \aap, 248, 485

\bibitem[Eggenberger, Udry, \& Mayor(2004)]{Egg:04::} 
Eggenberger, A., Udry, S., \& Mayor, M.\ 2004, \aap, 417, 353

\bibitem[Gilliland et al.(1992)]{Gil:92::} Gilliland, R.~L., 
Morris, S.~L., Weymann, R.~J., Ebbets, D.~C., \& Lindler, D.~J.\ 1992, 
\pasp, 104, 367

\bibitem[Graziani \& Black(1981)]{Gra:81::} Graziani, F.~\& 
Black, D.~C.\ 1981, \apj, 251, 337

\bibitem[Griffin \& Griffin(1973)]{Grif:73::} Griffin, R.~\& Griffin, R.\ 
1973, \mnras, 162, 243

\bibitem[Holman \& Wiegert(1999)]{Hol:99::} Holman, M.~J.~\& 
Wiegert, P.~A.\ 1999, \aj, 117, 621

\bibitem[Konacki \& Lane(2004)]{Kon:04::} Konacki, M.~\& Lane, 
B.~F.\ 2004, \apj, 610, 443

\bibitem[Konacki, Torres, Sasselov, \& Jha(2003a)]{Kon:03a::} 
Konacki, M., Torres, G., Sasselov, D.~D., \& Jha, S.\ 2003a, \apj, 597, 1076

\bibitem[Konacki, Torres, Jha, \& Sasselov(2003b)]{Kon:03b::} 
Konacki, M., Torres, G., Jha, S., \& Sasselov, D.~D.\ 2003b, \nat, 421, 507

\bibitem[Kurucz(1995)]{Kurucz:95::} Kurucz, R.~L. 1995, ASP
Conf.~Ser.~ 78: Astrophysical Applications of Powerful New Databases,
205

\bibitem[Lastennet \& Valls-Gabaud(2002)]{Las:02::} Lastennet, 
E.~\& Valls-Gabaud, D.\ 2002, \aap, 396, 551

\bibitem[Laughlin \& Chambers(2002)]{Lau:02::} Laughlin, G.~\& 
Chambers, J.~E.\ 2002, \aj, 124, 592

\bibitem[Lubow \& Artymowicz(2000)]{Lub:00::} Lubow, S.~H.~\& 
Artymowicz, P.\ 2000, Protostars and Planets IV, 731

\bibitem[Marcy, Butler, Fischer, \& Vogt(2003)]{Mar:03::} Marcy, 
G.~W., Butler, R.~P., Fischer, D.~A., \& Vogt, S.~S.\ 2003, ASP 
Conf.~Ser.~294: Scientific Frontiers in Research on Extrasolar Planets, 1

\bibitem[McCabe, Duch{\^ e}ne, \& Ghez(2003)]{Mcc:03::} McCabe, 
C., Duch{\^ e}ne, G., \& Ghez, A.~M.\ 2003, \apjl, 588, L113

\bibitem[Marcy \& Butler(1992)]{Mar:92::} Marcy, G.~W.~\& 
Butler, R.~P.\ 1992, \pasp, 104, 270

\bibitem[Mathieu, Ghez, Jensen, \& Simon(2000)]{Mat:00::} 
Mathieu, R.~D., Ghez, A.~M., Jensen, E.~L.~N., \& Simon, M.\ 2000, 
Protostars and Planets IV, 703

\bibitem[Marzari \& Scholl(2000)]{Mar:00::} Marzari, F.~\& 
Scholl, H.\ 2000, \apj, 543, 328

\bibitem[Mayor \& Mazeh(1987)]{May:87::} Mayor, M.~\& Mazeh, T.\ 
1987, \aap, 171, 157

\bibitem[Mermilliod, Rosvick, Duquennoy, \& 
Mayor(1992)]{Mer:92::} Mermilliod, J.-C., Rosvick, J.~M., 
Duquennoy, A., \& Mayor, M.\ 1992, \aap, 265, 513

\bibitem[Nadal, Ginestet, Carquillat, \& Pedoussaut(1979)]{Nad:79::} Nadal, R., 
Ginestet, N., Carquillat, J.-M., \& Pedoussaut, A.\ 1979, \aaps, 35, 203

\bibitem[Naef et al.(2004)]{Na:04::} Naef, D., Mayor, M., 
Beuzit, J.~L., Perrier, C., Queloz, D., Sivan, J.~P., \& Udry, S.\ 2004, 
\aap, 414, 351

\bibitem[Nelson(2000)]{Nel:00::} Nelson, A.~F.\ 2000, \apjl, 
537, L65

\bibitem[Nidever et al.(2002)]{Nid:02::} Nidever, D.~L., Marcy, 
G.~W., Butler, R.~P., Fischer, D.~A., \& Vogt, S.~S.\ 2002, \apjs, 141, 503

\bibitem[Pendleton \& Black(1983)]{Pen:83::} Pendleton, Y.~J.~\& 
Black, D.~C.\ 1983, \aj, 88, 1415

\bibitem[Pilat-Lohinger \& Dvorak(2002)]{Pil:02::} 
Pilat-Lohinger, E.~\& Dvorak, R.\ 2002, Celestial Mechanics and Dynamical 
Astronomy, 82, 143

\bibitem[Pilat-Lohinger, Funk, \& Dvorak(2003)]{Pil:03::} 
Pilat-Lohinger, E., Funk, B., \& Dvorak, R.\ 2003, \aap, 400, 1085 

\bibitem[Rabl \& Dvorak(1988)]{Rab:88::} Rabl, G.~\& Dvorak, R.\ 
1988, \aap, 191, 385

\bibitem[Rodriguez et al.(1998)]{Rod:98::} Rodriguez, L.~F., et 
al.\ 1998, \nat, 395, 355

\bibitem[Saar, Butler, \& Marcy(1998)]{Saa:98::} Saar, S.~H., 
Butler, R.~P., \& Marcy, G.~W.\ 1998, \apjl, 498, L153

\bibitem[Schneider(2004)]{Sch:04::} Schneider,~J.~\ 2004,
{The~Extrasolar~Planets~Encyclopaedia},
{\tt http://www.obspm.fr/encycl/encycl.html}

\bibitem[Simon et al.(1995)]{Sim:95::} Simon, M., et al.\ 1995, 
\apj, 443, 625

\bibitem[Quirrenbach(2001)]{Qui:01::} Quirrenbach, A.\ 2001,
\araa, 39, 353

\bibitem[Udry, Mayor, \& Queloz(1999)]{Ud:99::} Udry, S., 
Mayor, M., \& Queloz, D.\ 1999, ASP Conf.~Ser.~185: IAU Colloq.~170: 
Precise Stellar Radial Velocities, 367

\bibitem[Whitmire, Matese, Criswell, \& Mikkola(1998)]{Whi:98::} 
Whitmire, D.~P., Matese, J.~J., Criswell, L., \& Mikkola, S.\ 1998, Icarus, 
132, 196

\bibitem[Zucker \& Mazeh(1994)]{Zuc:94::} Zucker, S.~\& Mazeh, 
T.\ 1994, \apj, 420, 806

\end{thebibliography}
\end{document}